\documentclass[epsfig,12pt,onecolumn]{article}
\usepackage{amsfonts}
\usepackage{amssymb}
\usepackage{amsmath}
\usepackage{multicol}
\usepackage{graphicx}
\usepackage{float}
\usepackage{caption}

\input{tcilatex}
\makeatletter \@addtoreset{equation}{section}

\def \be{\begin{equation}}
\def \ee{\end{equation}}
\def \bea{\begin{eqnarray}}
\def \eea{\end{eqnarray}}

\newcommand{\nc}{\newcommand}
\nc{\al}{\alpha} \nc{\bib}{\bibitem} \nc{\la}{\lambda}
\nc{\C}{\mbox{\hspace{1.24mm}\rule{0.2mm}{2.5mm}\hspace{-2.7mm} C}}
\nc{\R}{\mbox{\hspace{.04mm}\rule{0.2mm}{2.8mm}\hspace{-1.5mm} R}}

\begin{document}

\title{A new constant behind the rotational velocity of galaxies}
\author{M. Bousder$^{1}$\thanks{%
mostafa.bousder@gmail.com} \\
%EndAName
$^{1}${\small LPHE-MS Laboratory, Department of physics,}\\
\ {\small Faculty of Science, Mohammed V University, Rabat, Morocco}}
\maketitle

\begin{abstract}
The present work is devoted to studying the dynamical evolution of galaxies
in scalar-Gauss-Bonnet$\ $gravity and its relationship with the MOND
paradigm. This study is useful for giving meaning to the presence of a new
gravitational constant. The stability of dark matter is strongly dependent
on matter density. We are interested in calculating the maximum rotational
velocity of galaxies. We show that rotating galaxies can be described by a
new parameter that depends both on the minimum value of scalar fields and on
the effective mass of this field. According to observational data, we have
shown that this parameter is a constant.

\textbf{Keywords:} Dark matter, Galaxies, Einstein-Gauss-Bonnet\ gravity.
\end{abstract}

\section{Introduction}

Recently, several models of extended or modified gravity theories were
proposed \cite{0A}, to explain the missing gravity problem \cite{0DM}, one
of the major problems in modern cosmology. According to Lovelock theorem
\cite{0L}, the Gauss-Bonnet (GB) gravity is\ introduced only in case $D>4$.
In four-dimensional spacetime, the GB term does not contribute to the
gravitational dynamics. Recently, there has been renewed interest in the GB
gravity, D. Glavan and C. Lin \cite{G1} proposed a novel 4-dimensional
Einstein-Gauss-Bonnet (EGB) gravity, which has attracted great attention.
Their idea is to multiply the GB term by $1/(D-4)$ before taking the limit.
This offers a new 4-dimensional gravitational theory with only two dynamical
degrees of freedom by consider the $D\longrightarrow 4$ limit of EGB gravity
in $D>4$ dimensions \cite{G2}, which is in contradiction with Lovelock
theorem. However, it was shown in several papers that perhaps the idea of
the limit $D\longrightarrow 4$ is not clearly defined. Several ideas have
been proposed to remedy this inconsistency and the absence of a proper
action \cite{F1,F2,F3,F4}.\textbf{\ }Although the EGB gravity is currently
debatable, the spherically symmetric black hole solution is still meaningful
and worthy of study \cite{N1}. There is a little work on the study of dark
matter in the context of EGB gravity \cite{NN2}. Note that the $R^{2}$
gravity models are an increasingly important area of research to study the
missing gravity problem \cite{1s}. Since GB contains an $R^{2}$ term, in
this case, we can propose that GB gravity generalize $R^{2}$ gravity. Khoury
and Weltman \cite{2s} proposed a new coupling that gives to the scalar field
a mass depending on the local density of matter. The modified Newtonian
dynamics (MOND) an effective theory paradigm proposed to explain the problem
of flat rotation curve of spiral galaxies. It constitutes an alternative to
the concept of dark matter \cite{13a}. MOND constitutes a modification to
Newtonian dynamics in the limit of low accelerations. This would mean that
MOND might emerge as an approximate consequence of some deeper physical
theory \cite{mil}. In the present paper, we consider a model of a scalar
field $\phi $ in the context of EGB\ gravity. This scalar field will
describe the dark matter. Thus this model very robustly leads to the maximum
values of rotational velocity of galaxies and dust for scalar field
potentials. \newline
The MOND model enables a broadening of the range of scales that are
theoretically well understood, from the kpc scales of galactic bars to the
Gpc scale of the local void and the Hubble tension \cite{MO1}. MOND can
account for the Hubble tension by means of outflow from a large local
supervoid, which has been observed and is known as the KBC void \cite{REF}.
While outflows from voids are expected in $\Lambda$CDM, structure formation
would be enhanced in MOND, allowing it to explain the formation of the KBC
void even though $\Lambda$CDM cannot \cite{MO3}. A number of theories of
gravity have studied dark matter in the regime of galaxies according to the
relativistic MOND theory \cite{MO4,MO5}. MOND can also account for the
massive high-redshift galaxy cluster collision known as El Gordo, which
contradicts $\Lambda$CDM at high significance \cite{MO6}\newline
A group of galaxies was studied using the MOND and the dark haloes, in view
of two suggested explanations for the discrepancy between the luminous mass
and the conventional dynamical mass of galaxies \cite{MO7,MO8,MO9}.\newline
This paper is organized as follows. In the next section, we introduce the
model of the scalar field $\phi $ minimally coupled to EGB gravity. Section
3 is devoted to analyzing the mass of the scalar field. In section 4, we
discuss the\textbf{\ }stability of EGB$\ $gravity. In section 5, we perform
analytic analyses of the rotation curve of the galaxies. The last section is
devoted to the conclusion.\newline

\section{Minimally coupled to EGB$\ $gravity}

Recently, there has been a renewed interest in the relationship between dark
matter and the scalaron mass \cite{K1,D0}. Consider now the
scalar-Gauss-Bonnet gravity in $4$-dimensions \cite{D1,D2}:%
\begin{equation}
S=\int d^{4}x\sqrt{-g}\left( \frac{M_{p}^{2}}{2}R+f\left( \phi \right)
\mathcal{G}-\frac{1}{2}g^{\mu \nu }\partial _{\mu }\phi \partial _{\nu }\phi
-V\left( \phi \right) \right) ,  \label{EG1}
\end{equation}%
where $M_{p}^{2}=\frac{c^{4}}{8\pi G}$, $R$ is the Ricci scalar, and $%
f\left( \phi \right) $ is the Gauss-Bonnet coupling function with dimensions
of $\left[ length\right] ^{2}$, that represent ultraviolet (UV) corrections
to Einstein theory. In the above equation $\left( \mu ,\nu \right) =\left(
0,1,2,3\right) $. We define the Gauss-Bonnet invariant as
\begin{equation}
\mathcal{G}\equiv R^{2}-4R_{\mu \nu }R^{\mu \nu }+R_{\mu \nu \rho \sigma
}R^{\mu \nu \rho \sigma }.  \label{EG2}
\end{equation}%
The variation with respect to the field $\phi $ gives us the equation of
motion for the scalaron field%
\begin{equation}
\square \phi -\partial _{\phi }V\left( \phi \right) +\mathcal{G}\partial
_{\phi }f\left( \phi \right) =0.  \label{EG4}
\end{equation}%
The variation of the action over the metric $g_{\mu \nu }$ simplified by the
Bianchi identity gives%
\begin{eqnarray}
0 &=&M_{p}^{2}\left( R^{\mu \nu }-\frac{1}{2}g^{\mu \nu }R\right) +\frac{1}{2%
}\partial ^{\mu }\phi \partial ^{\nu }\phi -\frac{1}{4}g^{\mu \nu }\partial
_{\rho }\phi \partial ^{\rho }\phi +\frac{1}{2}g^{\mu \nu }\left( -V\left(
\phi \right) +f\left( \phi \right) \mathcal{G}\right)  \label{EG5} \\
&&f\left( \phi \right) \left( -2RR^{\mu \nu }+R_{\text{ \ }\rho }^{\mu
}R^{\nu \rho }-2R^{\mu \rho \sigma \tau }R_{\text{ \ }\rho \sigma \tau
}^{\nu }+4R^{\mu \rho \sigma \tau }R_{\text{ \ }\rho \sigma \tau }^{\nu
}\right)  \notag \\
&&+\left( 2R\nabla ^{\mu }\nabla ^{\nu }-2g^{\mu \nu }R\nabla ^{2}-4R^{\nu
\rho }\nabla _{\rho }\nabla ^{\mu }-4R^{\mu \rho }\nabla _{\rho }\nabla
^{\nu }\right) f\left( \phi \right)  \notag \\
&&+4\left( \nabla ^{2}f\left( \phi \right) \right) R^{\mu \nu }+4g^{\mu \nu
}\left( \nabla _{\rho }\nabla _{\sigma }f\left( \phi \right) \right) R^{\rho
\sigma }-4\left( \nabla _{\rho }\nabla _{\sigma }f\left( \phi \right)
\right) R^{\mu \rho \nu \sigma }.  \notag
\end{eqnarray}%
The metric of a spatially flat homogeneous and isotropic universe in FLRW
model is given by:%
\begin{equation}
ds^{2}=-dt^{2}+a^{2}(t)\sum_{i=1}^{3}\left( dx^{i}\right) ^{2},  \label{EG6}
\end{equation}%
where $a(t)$ is a dimensionless scale factor, from which we define the Ricci
scalar $R$ and the GB invariant $\mathcal{G}$ in FLRW geometry as%
\begin{equation}
R=6\left( 2H^{2}+\dot{H}\right) \text{ \ \ \ \ }\mathcal{G}=24H^{2}\left(
\dot{H}+H^{2}\right) .  \label{RR}
\end{equation}%
We start by considering $\phi =\phi \left( t\right) $. So, Eq.(\ref{EG5}) is
written as \
\begin{equation}
\frac{1}{2}\dot{\phi}^{2}+24H^{3}f^{\prime }\left( \phi \right) \dot{\phi}%
+V\left( \phi \right) =3M_{p}^{2}H^{2},  \label{EG7}
\end{equation}%
where $\dot{\phi}\equiv \partial _{t}\phi $, $f^{\prime }\left( \phi \right)
\equiv \partial _{\phi }f\left( \phi \right) $, and $H\equiv \dot{a}/a$ is
the Hubble parameter. The scalar term vanishes if $\dot{\phi}=0$, leading to
$V\left( \phi =constant\right) =3M_{p}^{2}H^{2}$ (the cosmic critical
density). The solution Eq.(\ref{EG7}) is then evaluated in the form of an
energy equation. There is a dark matter sector with $\Omega _{DM}=\dot{\phi}%
^{2}/6M_{p}^{2}H^{2}$ and a dark energy sector with $\Omega _{DE}=V\left(
\phi \right) /3M_{p}^{2}H^{2}$. For interaction between these sectors, we
have $\Omega _{I}=24Hf^{\prime }\left( \phi \right) \dot{\phi}/3M_{p}^{2}$.\
The fraction of dark matter is $\Omega _{DM}=1-\Omega _{DE}-\Omega _{I}$. It
is possible, that for appropriate choices of the potential and the coupling
function, that competition between these terms leads to a minimum in the
effective potential. We consider in this work the dilatonic-type:%
\begin{equation}
V\left( \phi \right) =V_{0}e^{-k\phi }\text{, \ }f\left( \phi \right)
=f_{0}e^{+k\phi },  \label{EG8}
\end{equation}%
in this case, we refer to the scalaron as a dilaton field. Indeed, the
coupling function remains invariant under the simultaneous sign change $%
\left( k,\phi \right) \rightarrow \left( -k,-\phi \right) $. The equatios of
motion (Eq.\ref{EG4}) is invariant under the transformation $V\left( \phi
\right) \longleftrightarrow -\mathcal{G}f\left( \phi \right) $. In what
follows, we shall assume that the term$-\mathcal{G}f\left( \phi \right) $
represents the second term in the effective potential. In that case the
harmonic term dominates the potential, so one can approximate the equation
of motion to find a damped harmonic oscillation \cite{D1}.

\section{The scalaron mass}

Next, we study the mass of the scalar field which will describe the mass of
dark matter.\ In order to find an explicit expression for the scalaron mass,
we need to consider an effective potential in the equation of motion (Eq.\ref%
{EG4}), which can be written as the Klein Gordon equation in the FLRW metric
as:%
\begin{equation}
\ddot{\phi}+3H\dot{\phi}+\partial _{\phi }V\left( \phi \right) -\mathcal{G}%
\partial _{\phi }f\left( \phi \right) =0.  \label{c1}
\end{equation}%
\ Let us now use the above expressions to examine the evolution of the
scalaron. Observe that the term $V\left( \phi \right) -\mathcal{G}f\left(
\phi \right) $ acts as the effective potential for the perturbations. The
effective potential \cite{AZ1,AZ2} that includes the GB term can be written
as:
\begin{equation}
V_{eff}=V\left( \phi \right) -\mathcal{G}f\left( \phi \right) .  \label{c2}
\end{equation}%
We notice that the effective potential of the scalaron includes the
Gauss-Bonnet coupling and the Gauss-Bonnet invariant. In other words, the
Gauss-Bonnet term affects the potential structure of the scalaron, so the
scalaron mass depends on the matter contribution. The particles of the field
$\phi $ come from the fluctuation around the minimum of the effective
potential $V_{eff}(\phi )$. Using (\ref{EG8}), the second derivative of the
effective potential Using $\phi $ is
\begin{equation}
\frac{\partial ^{2}}{\partial \phi ^{2}}V_{eff}=k^{2}V_{0}e^{-k\phi }-k^{2}%
\mathcal{G}f_{0}e^{+k\phi }.  \label{c3}
\end{equation}%
As before the existence of a minimum in the effective potential requires $%
\mathcal{G}\neq 0$. More precisely, we define the effective mass as a
function of the field $\phi $%
\begin{equation}
m_{eff}^{2}=\left. \frac{\partial ^{2}}{\partial \phi ^{2}}V_{eff}(R,\phi
)\right\vert _{\phi =\phi _{\min }},  \label{c4}
\end{equation}%
where $\phi _{\min }$\ is the minimum value of the scalaron $\phi $. The
effective mass is then given by%
\begin{equation}
m_{eff}=k\sqrt{V_{0}e^{-k\phi _{\min }}-\mathcal{G}f_{0}e^{+k\phi _{\min }}}.
\label{c5}
\end{equation}%
When the minimum of the scalaron is very small, the solution could be given
by $m_{eff}=k\sqrt{V_{0}-\mathcal{G}f_{0}}$. The domain of validity of the
fluid description of dark matter is when $m_{eff}$ is real. One finds that $%
\mathcal{G}$ has the bound%
\begin{equation}
\mathcal{G}\leq \frac{V_{0}}{f_{0}}e^{-2k\phi _{\min }}.  \label{c6}
\end{equation}%
If the condition is satisfied everywhere in spacetime, then the adiabatic
approximation will be good everywhere. In particular, considering the
maximum value of $\mathcal{G}$ is:%
\begin{equation}
\mathcal{G}_{\max }=\frac{V_{0}}{f_{0}}e^{-2k\phi _{\min }}.  \label{c7}
\end{equation}%
For $\mathcal{G=G}_{\max }$, the scalaron mass will be zero. The above
inequality Eq.(\ref{c7}) can be written as%
\begin{equation}
q\geq \frac{-V_{0}}{24H^{4}f_{0}}e^{-2k\phi _{\min }},  \label{c8}
\end{equation}%
where $q\equiv -1-\dot{H}/H^{2}=-a\ddot{a}/\dot{a}^{2}$\ is the deceleration
parameter. We mention that according to Eq.(\ref{c8}), even if $\mathcal{G}$
$=\mathcal{G}_{\max }$, the asymptotic value of $q$ depends on the model
parameters $V_{0}$ and $f_{0}$. Furthermore the isolated points where $%
q=q_{\min }$ constitute a set of measure zero and do not contribute to $%
m_{eff}$.

\section{Dark matter stability analysis}

Let us describe the effective potential in an environment that surrounds the
matter. The GB invariant can be greatly simplified to the matter density $%
\rho $ \cite{AZ1}, giving\newline
\begin{equation}
\mathcal{G}=\rho .  \label{DM1}
\end{equation}%
Therefore, in order to satisfy Eq.(\ref{c6}) one must set that $V_{0}$ and $%
f_{0}$ have the same sign. From Eqs.(\ref{EG8},\ref{c2},\ref{c7},\ref{DM1})
and we adopt that $\mathcal{G}_{\max }=\rho_{\max }$ we finally obtain
\begin{equation}
V_{eff}=f_{0}e^{-k\phi }\left( \rho _{\max }e^{2k\phi _{\min }}-\rho
e^{2k\phi }\right) ,  \label{DM2}
\end{equation}%
this expression is comparable to that found by \cite{AZ1}. At low $k\phi $,
the effective potential can be approximated by $V_{0eff}$. In the high $%
k\phi $ regime,\ $V_{eff}$ to $V_{0}\exp (k\phi )$. Our discussion of
stability in the following sections will be valid for $m_{eff}^{2}\left(
\phi \right) >0$. We can then express the mass $m_{eff}$ as a function of
the matter energy density%
\begin{equation}
m_{eff}^{2}=k^{2}f_{0}e^{-k\phi _{\min }}\left( \rho _{\max }-\rho \right) ,
\label{DM3}
\end{equation}%
where we have taken into account that $\rho _{\max }=\frac{V_{0}}{f_{0}}%
e^{-2k\phi _{\min }}$. The mass of small fluctuations is an increasing
function of the matter energy density. At low density, the effective mass
can be approximated by $k^{2}V_{0}e^{-3k\phi _{\min }}$. In the high-density
region,\ $m_{eff}$ to $0$, corresponding to the dark matter halos around
galaxies. To be more precise, dark matter is very important in the
extremities of the galaxy when $\rho \sim 0$. Apart from the evolution of
the scalaron that was extracted above, the other important consequence is
that when the matter energy density becomes maximal, the mass of this field
will be zero. The effective pressure and density are $p_{eff}=-V\left( \phi
\right) +f\left( \phi \right) \rho $ and $\rho _{eff}=V\left( \phi \right) $%
, respectively. The effective equation of state for this system is given by:%
\begin{equation}
\omega _{eff}\equiv \frac{p_{eff}}{\rho _{eff}}=-1+\frac{\rho }{\rho _{\max }%
}e^{2k\left( \phi -\phi _{\min }\right) }.  \label{DM4}
\end{equation}%
We mention that the $\omega _{eff}$ lies in the density $\rho $ and the
scalaron $\phi $. If $V_{0}>0$, the pressure for this system will be
negative. The system will be stable if
\begin{equation}
c_{eff}^{-2}\equiv \frac{d\rho _{eff}}{dp_{eff}}>0,  \label{DM5}
\end{equation}%
where $c_{eff}$ is the effective sound speed. This equation gives a simple
prescription for computing when a given theory will be stable. We calculate $%
c_{eff}$ for a fixed density, allowing us to find that $\omega
_{eff}=1/c_{eff}^{2}$. We mention that for $\omega _{eff}=0$, we find that $%
\rho _{\max }=\rho e^{2k\left( \phi -\phi _{\min }\right) }$. Even if $\phi
\sim \phi _{\min }$, one obtains the asymptotic value $\rho \simeq \rho
_{\max }$. Additionally, to assess possible behaviors of the sound speed
squared and $\omega _{eff}$, we remark that the stable regime arises in the
case%
\begin{equation}
\rho _{\max }<\rho e^{2k\left( \phi -\phi _{\min }\right) }.  \label{DM6}
\end{equation}%
In this expression, $k$ should be positive so that the theory will be stable
(since $\phi >\phi _{\min }$). If $k<0$, the mass $m_{eff}$ correspond to
the tachyonic instability. Inserting $\rho <\rho _{\max }$ into Eq.(\ref{DM6}%
) we get the result that $1<\frac{\rho _{eff}}{\rho }<e^{2k\left( \phi -\phi
_{\min }\right) }$, it is necessary that $k>0$. In the regime of large
densities, we have $\phi >\phi _{\min }$. If $\phi \gg \phi _{\min }$, the
effective sound speed can be greatly simplified, giving $c_{eff}^{2}>0$. In
summary, this analysis shows that the EGB dark matter lies in the stable
regime. Since the right hand side involves exponential factors, we expect
that the field $\phi $ to evolve by a logarithmic function of the matter
density. \newline

\section{Rotation curves from MOND and the relation to EGB}

It is well known that the orbital velocities $v$ of planets in planetary
systems and moons orbiting planets decline with distance according to
Kepler's third law \cite{80} $v^{2}=\frac{GM}{r},$ with $G\approx 6,67\times
10^{-11}m^{3}Kg^{-1}s^{-2}$. In contrast, stars revolve around their
galaxy's center at equal or increasing speed over a large range of distances
\cite{9}. The rotational speeds of stars inside the galaxies do not follow
the rules found in smaller orbital systems. A solution to this conundrum is
to suppose the existence of dark matter and to tune its distribution from
the galaxy's center out to its halo based on the observed kinematics \cite%
{10}. The effective potential can be used to determine the orbits of planets
\cite{101}, also in the cosmological evolution analysis of the chameleon
field, allowing the detection of dark energy in orbit \cite{102}.\ According
to the MOND theory \cite{13a}, the rotational velocity of stars around a
galaxy at large distances is:%
\begin{equation}
v_{0}^{4}=GMa_{0}.  \label{1}
\end{equation}%
where $a_{0}\approx 1,2\times 10^{-10}ms^{-2}$ and $M$ is the total mass of
a galaxy. It is treated as a point mass at its centre, providing a crude
approximation for a star in the outer regions of a galaxy. The Eq.(\ref{1})
predicts that the rotational velocity is constant out to an infinite range
and that the rotational velocity doesn't depend on a distance scale, but on
the magnitude of the acceleration $a_{0}$. We suppose that the constant $%
a_{0}$ ni more constant, after seeing the form of galaxy rotation curves in
MOND \cite{mil}. We start by changing%
\begin{equation}
v_{0}^{4}\rightarrow GM\left( \frac{GM}{f_{0}}\right) ,  \label{2}
\end{equation}%
where $f_{0}$ has dimensions of $\left[ length\right] ^{2}$. We can write
that the rotational velocity of a galaxy is:
\begin{equation}
v^{2}\left( r\right) =\frac{GMk}{m_{eff}}e^{-\frac{k\phi _{\min }}{2}}\sqrt{%
\rho _{\max }-\rho \left( r\right) },  \label{3}
\end{equation}%
where $\rho $ is the local matter energy density, $\rho _{\max }$ is the
maximum density of the galaxy and $M$ is its mass. We consider a disk of
radius $r$ with its center at the galactic center. This modification based
on EGB gravity instead of Newton's gravity (as in the case of MOND theory),
will later introduce a relativistic term to the potential. Moreover, the
study of relativistic treatment within the framework of MOND theory is
treated in \cite{MO1, MO2}. The choice of this potential has several
advantages, since it generalizes the Newtonian potential and it has a
relation with the potential of MOND theory \cite{13a}. The potential above
has a relativistic aspect which is related to the parameters of EGB gravity.
The term $\sqrt{1-\frac{\rho \left( r\right) }{\rho _{\max }}}$ represents
the relativistic part of this potential. The description of the rotation of
galaxies in the relativistic EGB gravity is better compared to the Newtonian
frame, and gives a complete dynamic in space-time. On the other hand, the
MOND theory remains incomplete since it has a lack of relativistic treatment
of the rotation of galaxies. Note that in the edges of the galaxy we take a
low matter energy density $\rho \left( r_{\max }\right) \sim 0$, yielding:%
\newline
\begin{equation}
V_{\max }\sim \left( GM\right) ^{1/2}\left( \rho _{\max }\right)
^{1/4}\theta _{0},  \label{x3}
\end{equation}%
where $\theta _{0}=\left( ke^{-\frac{k\phi _{\min }}{2}}/m_{eff}\right)
^{1/2}$. The parameter $\theta _{0}$ strongly affects the behavior of the
velocity $V_{\max }$. The galaxy's rotation curves remain almost constant or
increasing at the edges of galaxies \cite{9,11,13}. To describe the maximum
rotational velocity of a galaxy, we use the speed $V_{\max }$ Eq.(\ref{x3}).
which corresponds to an almost constant rotation if $\theta _{0}$ is fixed.
In what follows, we will draw the curve of $\theta _{0}$ according to the
observations. To determine the value of $\theta _{0}$, we study the maximum
values of rotational velocity $V_{\max }$ of galaxies according to their
density $\rho $ and the masse $M_{in}$ of galaxies. Using the observation
results, we can determine the values of $\theta _{0}$ \cite{13d,13h,13e,13f}:%
\begin{equation}
\theta _{0}\approx 8.167\times 10^{-6}\times V_{\max }\times \rho _{\max }^{-%
\frac{1}{4}}\times M^{-\frac{1}{2}}.
\end{equation}%
To find the value of $\theta _{0}$ we have to calculate $\left( M,\rho
_{\max },V_{\max }\right) $ for some galaxies in the tables of the Appendix.
\begin{figure}[H]
\centering
\includegraphics[width=11cm]{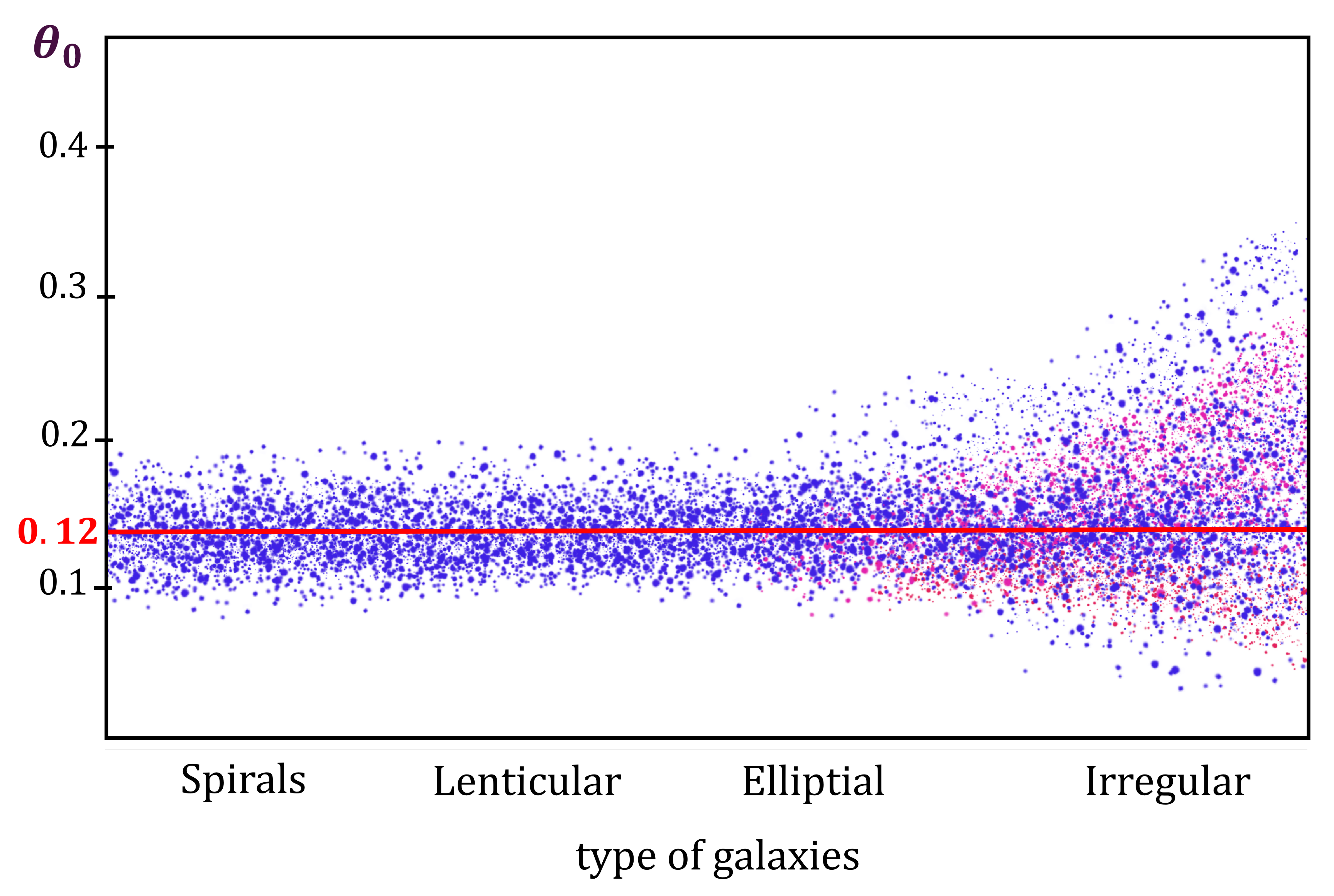}
\caption{$\protect\theta _{0}$ vs type of galaxies. These points are
obtained from observations \protect\cite{9}. Notice that the values of $%
\protect\theta _{0}$ are almost constant ($\protect\theta _{0}\approx 0.12$).
}
\label{graph1}
\end{figure}
According to Figure (\ref{graph1}), the parameter $\theta _{0}$ permit us to
determine the rotational velocity of galaxies. The values of $\theta _{0}$
are very close under wide range of conditions, but there are exceptions like
dwarf spheroidal galaxies.

\section{Summary}

We have presented a study that is designed to describe dark matter in
Einstein-Gauss-Bonnet (EGB)\ gravity coupled with scalar fields. We also
analyzed the effects of the GB invariant on the evolution of effective
potential. Also, the stability of dark matter was discussed in the context
of this model. We have presented an explicit procedure to construct the
effective equation of state describing the scalar fields. The potential
describes the rotation of galaxies by two parameters $k$ and $m_{eff}$
introduced by EGB gravity. Indeed, these two parameters describe the dark
matter hidden in galaxies. We have compared our results with observations.
We have used the EGB gravity and MOND theory, in relationship with the field
$\phi $ to provide a qualitative description of modified gravity. We found
an expression of for how the rotation curve of a galaxy depends on the
parameter $\theta _{0}$. We obtained a graph (Figure \ref{graph1}) of $%
\theta _{0}$ using the observational data (Appendix) which corresponds
exactly to the evolution of the rotation of galaxies. We have shown that $%
\theta _{0}$ is an essential parameter, which is constant for spirals,
lenticular and elliptical galaxies, but is no longer a constant for
irregular and dwarf spheroidal galaxies. This shows that the parameter $%
\theta _{0}$ plays an important role in describing both the rotation and the
type of galaxies. Future work will have to test if this model also
corresponds to other observations like the CMB.

\section{Appendix}

\begin{table}[H]
\begin{equation*}
\begin{tabular}{ccccc}
\hline
Galaxy & $\rho (\times 10^{-22}kg/m^{2})$ & $M_{in}(\times 10^{10}M_{\odot
}) $ & $V_{\max }(km/s)$ & $\theta _{0}$ \\ \hline
Milky Way & $13,10$ & $12,10$ & $220$ & $0,086$ \\ \hline
NGC 7331 & $8,70$ & $14,70$ & $268.1$ & $0,105$ \\ \hline
NGC 4826 & $45,00$ & $1,90$ & $180.2$ & $0,130$ \\ \hline
NGC 6503 & $18,40$ & $0,958$ & $121$ & $0,154$ \\ \hline
NGC 7793 & $12,00$ & $0,88$ & $117.9$ & $0,174$ \\ \hline
UGC 2885 & $15,30$ & $11,70$ & $300$ & $0,114$ \\ \hline
NGC 253 & $10,00$ & $4,30$ & $229$ & $0,160$ \\ \hline
NGC 925 & $8,60$ & $2,00$ & $113$ & $0,120$ \\ \hline
NGC 2403 & $5,20$ & $2,90$ & $143.9$ & $0,144$ \\ \hline
NGC 2841 & $8,02$ & $17$ & $326$ & $0,121$ \\ \hline
NGC 2903 & $6,30$ & $6,70$ & $215.5$ & $0,135$ \\ \hline
NGC 3198 & $3,26$ & $6,00$ & $160$ & $0,125$ \\ \hline
NGC 5585 & $10,1$ & $0,59$ & $92$ & $0,173$ \\ \hline
NGC 4321 & $8,80$ & $16,8$ & $270$ & $0,098$ \\ \hline
\end{tabular}%
\end{equation*}%
\caption{Large spirals galaxies}
\label{table:1}
\end{table}
\begin{table}[H]
\begin{equation*}
\begin{tabular}{ccccc}
\hline
Galaxy & $\rho (\times 10^{-22}kg/m^{2})$ & $M_{in}(\times 10^{10}M_{\odot
}) $ & $V_{\max }(km/s)$ & $\theta _{0}$ \\ \hline
NGC 4303 & $6,81$ & $3,68$ & $150$ & $0,124$ \\ \hline
NGC 5055 & $5,21$ & $7,07$ & $215$ & $0,138$ \\ \hline
NGC 4736 & $14,00$ & $1,77$ & $198.3$ & $0,198$ \\ \hline
NGC 5194 & $1,00$ & $4,00$ & $232$ & $0,299$ \\ \hline
NGC 4548 & $3,20$ & $3,80$ & $290$ & $0,287$ \\ \hline
\end{tabular}%
\end{equation*}%
\caption{Messier Spirals}
\label{table:2}
\end{table}
\begin{table}[H]
\begin{equation*}
\begin{tabular}{ccccc}
\hline
Galaxy & $\rho (\times 10^{-22}kg/m^{2})$ & $M_{in}(\times 10^{10}M_{\odot
}) $ & $V_{\max }(km/s)$ & $\theta _{0}$ \\ \hline
UGC 3993 & $3,10$ & $17.8$ & $300$ & $0,138$ \\ \hline
NGC 7286 & $4,60$ & $0.59$ & $98$ & $0,224$ \\ \hline
NGC 2768 & $10,00$ & $1.98$ & $260$ & $0,268$ \\ \hline
NGC 3379 & $0,90$ & $1.10$ & $60$ & $0,151$ \\ \hline
NGC 2434 & $1,00$ & $5.00$ & $231$ & $0,266$ \\ \hline
NGC 4431 & $13,00$ & $0.30$ & $78$ & $0,193$ \\ \hline
\end{tabular}%
\end{equation*}%
\caption{Lenticular and Elliptical Galaxies}
\label{table:3}
\end{table}
\begin{table}[H]
\begin{equation*}
\begin{tabular}{ccccc}
\hline
Galaxy & $\rho (\times 10^{-22}kg/m^{2})$ & $M_{in}(\times 10^{10}M_{\odot
}) $ & $V_{\max }(km/s)$ & $\theta _{0}$ \\ \hline
WLM (DDO 221) & $0,92$ & $0,00863$ & $19$ & $0,539$ \\ \hline
M81dWb & $5,00$ & $0,007$ & $28,5$ & $0,588$ \\ \hline
Holmberg II & $3,64$ & $0,0428$ & $34$ & $0,307$ \\ \hline
NGC 3109 & $8,00$ & $0,0299$ & $67$ & $0,605$ \\ \hline
NGC 4789a & $93,00$ & $0,0188$ & $50$ & $0,303$ \\ \hline
NGC 3034 & $22,00$ & $1,00$ & $137$ & $0,163$ \\ \hline
\end{tabular}%
\end{equation*}%
\caption{irregular dwarf galaxies}
\label{table:4}
\end{table}
\begin{table}[H]
\begin{equation*}
\begin{tabular}{ccccc}
\hline
Galaxy & $\rho (\times 10^{-22}kg/m^{2})$ & $M_{in}(M_{\odot })$ & $V_{\max
}(km/s)$ & $\theta _{0}$ \\ \hline
Carina & $6,50$ & $3.38\times 10^{6}$ & $8,5$ & $0,007$ \\ \hline
Leo I & $13,60$ & $7.74\times 10^{6}$ & $12,5$ & $0,006$ \\ \hline
Draco & $7,40$ & $3.40\times 10^{6}$ & $12$ & $0,01$ \\ \hline
Fornax & $0,373$ & $12.40\times 10^{6}$ & $11,5$ & $0,01$ \\ \hline
\end{tabular}%
\end{equation*}%
\caption{dwarf spheroidal galaxies (dSphs)}
\label{table:5}
\end{table}

\end{document}